\documentstyle[aps,epsfig,floats]{revtex}
\draft
\input epsf
\newcommand{\lbfig}[1]{\refstepcounter{fig} \label{#1} }
\newcounter{fig}
\newcommand{\gsim}{\lower.7ex\hbox{$\;\stackrel{\textstyle>}{\sim}\;$}}
\newcommand{\lsim}{\lower.7ex\hbox{$\;\stackrel{\textstyle<}{\sim}\;$}}
\newcommand{\bm}[1]{{\mbox{\boldmath $#1$}}}
\def\tgf{t_{\rm gf}}
\def\Bgf{B_{\rm gf}}
\def\tdec{t_{\rm dec}}

\def\Bleft{\Big[}
\def\Bright{\Big]}

\begin{document}
%
%two columns:
\twocolumn[\hsize\textwidth\columnwidth\hsize\csname
@twocolumnfalse\endcsname
\vspace*{-15mm}
\begin{flushright}
\baselineskip=14pt
{ DAMTP-1999-39}\\
{ {\it astro-ph/9904022\/}}
\end{flushright}\vspace*{-1mm}

\title{Relaxing the Bounds on Primordial Magnetic Seed Fields}
\author{Anne-Christine Davis\footnotemark, Matthew
Lilley\footnotemark\
 and Ola
T\"ornkvist\footnotemark}
\address{Department of Applied
Mathematics and Theoretical Physics, University of
Cambridge,\\ Silver Street, Cambridge CB3~9EW, United Kingdom }
\date{March 27, 1999}
\maketitle
\begin{abstract}We
point out that the lower bound on the
primordial magnetic field required to seed the galactic dynamo
is significantly relaxed in an open universe or in a
universe with
a positive cosmological constant.
It is shown that, for reasonable cosmological parameters, primordial seed
fields of strength $10^{-30}$ Gauss or less at the time of
galaxy formation could explain observed galactic
magnetic fields.  As a consequence,
mechanisms of
primordial magnetic seed-field
generation that have previously been ruled out
%, on the grounds of
%giving too small strength or correlation length,
could well be viable.
We also comment on the
implications of the observation of micro-Gauss magnetic fields in
galaxies at high redshift.
\end{abstract}
%
%two columns:
]
\footnotetext{~\vspace{-.8cm}~}
\footnotetext{\footnotemark[1]
Electronic address: {\tt A.C.Davis@damtp.cam.ac.uk}}
\footnotetext{\footnotemark[2]
Electronic address: {\tt M.J.Lilley@damtp.cam.ac.uk}}
\footnotetext{\footnotemark[3]
Electronic address: {\tt O.Tornkvist@damtp.cam.ac.uk}}
%
%2345678901234567890123456789012345678901234567890123456789012345678901234
\renewcommand{\thefootnote}{\arabic{footnote}}
Magnetic fields pervade most astrophysical systems \cite{obsreview},
but their origin is
unknown.
Spiral galaxies are observed to possess large-scale magnetic fields
with strength of the order of \mbox{$10^{-6}$ G} and direction
aligned with the rotational motion.
A plausible explanation
is
that galactic magnetic fields result from the exponential amplification
of an initially weak
seed field by a mean-field dynamo \cite{dynamo,moffatt}.
Many proposals have been put forward regarding the origin of
such a
seed field.
One suggestion is that it might arise spontaneously
from non-parallel gradients of pressure and charge-density
during galaxy formation
\cite{Kulsrud}.
A wider range
of possibilities
is offered
if the seed field is of primordial origin.
This category includes
cosmological magnetic fields
\cite{field} as well as magnetic fields created by
any of a number of
early-universe
particle-physics mechanisms \cite{thewholebunchincludingOlasOriginpaper}
such as collisions of bubbles in a first-order phase transition
\cite{KibVil} or false-vacuum inflation \cite{turner,davis}.

The seed-field strength required at the time
of completed galaxy formation ($\tgf$)
 for a galactic dynamo
to produce
the present magnetic field strength
 \mbox{$B_0\sim 10^{-6}$ G}
is usually quoted in the range
$\sim 10^{-23}$ -- \mbox{$10^{-19}$ G}.
Such lower bounds are obtained
by considering the
dynamo amplification
in a
flat universe with zero cosmological constant for
``typical'' values of the parameters of the
$\alpha\omega$--dynamo.
The seed field must also be
coherent on a scale at least as large as the size of the
largest
turbulent eddy, $\sim 100$ pc
\cite{dynamo}.
Most proposed
models of primordial seed-field generation fail to meet these
requirements
as formulated above.

In this paper we address the issue in the light of recent developments
in cosmology. Observations of distant type-IA supernovae \cite{supernovae}
and of anisotropies in the cosmic microwave background (CMB) \cite{CMB} in
combination have made it
increasingly likely
that the universe
is less dense
than the critical density
and has a positive cosmological constant $\Lambda$.
Most previous studies of magnetic fields have assumed a $\Lambda=0$
universe with critical matter density.

We shall recalculate the constraints on primordial seed fields for
general Friedmann universes with matter density parameter $\Omega_0$
and vacuum energy density parameter $\lambda_0 \equiv \Lambda/(3
H_0^2)$ such that \mbox{$\Omega_0+\lambda_0 \leq 1$} (the subscript
$0$ here indicates quantities at present time and $H_0$ is the Hubble
parameter).  In addition to finding revised bounds on the seed field
at time $\tgf$ we shall
trace the evolution of the magnetic field back to
the time of radiation decoupling, $\tdec$.
Prior to decoupling the evolution of the magnetic field
proceeds via complicated plasma processes and
depends on
the field's
initial strength and
correlation length
\cite{kostas}.
After decoupling there is
sufficient residual
ionisation for the magnetic field to be frozen into the
plasma; the evolution is simple and independent of the mechanism of
generation.
Thus $\tdec$ is a natural epoch for imposing bounds on primordial magnetic
fields.

We begin by considering the $\alpha\omega$--dynamo \cite{dynamo,moffatt}
which is a well-studied model of amplification of magnetic fields.
It is powered by the differential rotation of the
galaxy in combination with the small-scale turbulent motion of ionised
gas. By
separating the magnetic field into a large-scale mean field $\bm{B}$
and a random, turbulent field $\bm{b}$ one obtains a system of equations
with exponentially growing
solutions $B_\varphi\propto e^{\Gamma t}$ for the azimuthal
component $B_\varphi$
of the mean field
in the plane of the disc. The dynamo amplification rate
$\Gamma$ appears as an eigenvalue that must be determined numerically.

Unfortunately the value of $\Gamma$ is rather sensitive to the parameters
of the dynamo model \cite{RST},
which include the root-mean-square
velocity and magnetic diffusivity of the turbulent plasma
as well as the angular-velocity
gradient $r\,d\omega(r)/dr$.
For reasonable estimates of these quantities, one obtains $\Gamma^{-1}$ in
the range $0.2 < \Gamma^{-1} <0.8$ [Gyrs].
Because of the exponential growth, such an uncertainty
quickly translates into an uncertainty of
many orders of magnitude in the
total amplification, which may or may not
rule out various seed-field mechanisms.
The point of this paper is not to linger on these
uncertainties
but rather to emphasise the tremendous increase in amplification that
will occur
in an open universe or in one with a positive cosmological constant {\em
for any value
of\/} $\Gamma$.
We shall present results for the two values
that appear most frequently in the literature, $\Gamma^{-1}=0.3$ Gyrs
\cite{RST} and
$\Gamma^{-1}=0.5$ Gyrs \cite{dynamo}.

Any magnetic seed field is exponentially amplified until
it reaches the equipartition energy ($B\sim$ few $\mu$G) when further
growth is suppressed by dynamical back reaction of the Maxwell
stresses on the turbulence. Assuming that the dynamo mechanism begins
to operate
around the time of completed galaxy formation $\tgf$,
the lower bound on the strength of the
seed field at this epoch is given by
\begin{equation}
\label{tgfbound}
\Bgf \geq B_0 e^{-\Gamma(t_0 - \tgf)}~,
\end{equation}
where $t_0$ is the age of the universe.
Expressions for $t_0$ in different cosmologies
are found in e.g.\ Refs. \cite{kolb} and \cite{lahav}. In particular,
for a given value of $H_0$, open
universes and universes with a positive cosmological constant are
significantly older than the $\Omega=1$ Einstein-de Sitter universe.

The time of galaxy formation $\tgf$ can be
estimated from a spherical collapse model. As we shall show presently,
galaxies of a given average density $\bar{\rho}_{\rm gal}$
have collapsed at
approximately the same time after the Big Bang for all
realistic cosmological models. Galaxies in an open or $\Lambda>0$
universe are therefore older, giving the dynamo mechanism more time
$t_0 - \tgf$ to operate. Consequently, a much smaller magnetic
field $\Bgf$ can
seed the dynamo and still give the observed micro-Gauss field $B_0$.

The spherical collapse model \cite{peebles,padmanabhan}
describes the non-linear collapse of a bounded spherical region with
average local density $\bar{\rho}_{\rm i}$
larger than the critical density at some
initial time $t_{\rm i}$ in the matter-dominated era. This overdensity
causes the sphere to break away from the Hubble expansion, reach
a maximal (turn-around) radius $r_{\rm m}$, and
eventually collapse to form a gravitationally-bound system.
The general equation of motion for a shell of radius $r$ enclosing mass
$M$ is \cite{lahav}
\begin{equation} \label{lambdanewton}
{1 \over 2} \left({d r \over d t}\right)^2 - {GM \over r} - {1 \over
  6}\Lambda r^2 = E~,
\end{equation}
where $E$ is a constant.
This equation can be separated to yield
\begin{equation} \label{lambda_de}
dt =  \pm {r_{\rm m}^{3 /2} \over \sqrt{2 G M}}
 { \sqrt{x} dx \over \sqrt{1 -
      (1 + \beta)x + \beta x^3}}~,
\end{equation}
where $x=r/r_{\rm m}$ and
$\beta=\Lambda r_{\rm m}^3/(6GM)$. The exact solution can be expressed
in terms of incomplete elliptic integrals \cite{edwards}, but we
choose instead to expand the
right-hand side of Eq.~(\ref{lambda_de})
in $\beta$, obtaining
the more convenient parametric solution
\begin{eqnarray}
\label{parametric}
r=  {r_{\rm m} \over 2} \left(1 - \cos \theta \right)~,\quad\quad\quad
\quad\quad\\*
t+T = {1 \over \sqrt{GM}} \left( {r_{\rm m} \over 2}
\right)^{3/2}\Bleft(\theta - \sin\theta)
\quad\quad\quad\nonumber\\*
\label{parametrict}
+\frac{\beta}{96}
(66\,\theta - 93\sin\theta + 15\sin 2\theta - \sin 3\theta) +
{\cal O}(\beta^2)\Bright~,
\end{eqnarray}
where $T$ is a constant which
can be neglected \cite{padmanabhan}.
We see that turn-around occurs at a time $t_{\rm m}$ corresponding to
$\theta=\pi$.
As the spherical region recollapses for $t>t_{\rm m}$,
random non-radial particle
velocities become important; the simple collapse model breaks down
and the collapse is halted at a final
radius $r_{\rm vir}$ given by the virial theorem.
For universes with zero cosmological constant $r_{\rm vir}=r_{\rm
m}/2$.
For $0\leq\beta<1/2$ (which is
required for collapse to occur) Lahav {\em et al.\/} \cite{lahav}
showed that $1/2 \geq r_{\rm vir} / r_{\rm m} > 0.366$ and also
obtained the
approximate relation
\begin{equation} \label{r/rm}
{r_{\rm vir} \over r_{\rm m}} = {1 - \beta  \over 2 - \beta}~.
\end{equation}
We can
estimate $\beta$ for a galaxy: Taking $0\leq\lambda_0
< 1$, $M = 10^{11} M_{\odot} \approx 2 \times
10^{45} {\rm g}$ and $r_{\rm vir} \lsim 15$ kpc
we get
\begin{equation}
\label{betabound}
0\leq
\beta  < { H_0^2 \over  2 G M}{r_{\rm vir}^3 \over \left(0.366\right)^3}
\sim 3 \times 10^{-5}.
\end{equation}
The small value of $\beta$ signifies that the vacuum energy density plays a
negligible role compared to the matter density in the collapse of
objects as small and dense as galaxies. From
Eq.~(\ref{r/rm})
it  follows that
we can set
$r_{\rm vir}=r_{\rm m}/2$ for all realistic values of the cosmological
constant. Moreover, we can neglect the $\beta$-dependent terms in
Eq.~(\ref{parametrict}).

It is generally assumed \cite{padmanabhan} that gravitational collapse
is complete at the
time $t_{\rm vir}>t_{\rm m}$ when $r$ approaches zero
in Eq.~(\ref{parametric}),
corresponding to $\theta\approx 2\pi$.
\footnote{The
naive
estimate, that collapse is complete
  when the radius $r$ in Eq.~(\ref{parametric}) reaches
  the virial radius $r_{\rm vir}$ (corresponding to $\theta=3\pi/2$),
is unrealistic as the radius
decreases more slowly during virialisation than the
spherical collapse model would imply.} This assumption is supported by
N-body simulations
and, because of the small
value of $\beta$, remains valid in any realistic
Friedmann cosmology. From Eq.~(\ref{parametrict}) we then have
$t_{\rm vir}\approx 2 t_{\rm m}$ as well as $t_{\rm m}^2=
3\pi/(32G\bar{\rho}_{\rm m})$, where $\bar{\rho}_{\rm m}\equiv 3M/(4\pi
r_{\rm m}^3)$ is the average density of the spherical region at
turn-around. It follows from $r_{\rm vir}=r_{\rm m}/2$ that
$\bar{\rho}_{\rm gal}=8 \bar{\rho}_{\rm m}$. Finally, with
$t_{\rm gf}=t_{\rm vir}$, all these relations combine to give
\begin{equation} \label{rhobar}
\bar{\rho}_{\rm gal} = {3 \pi \over G t_{\rm gf}^2}~.
\end{equation}

%%%%%%%%%%%%%%%%%%%%%%%%%%%%%%%%%%%%%%%%%%%%%%%%%%%%%%%%%%%%%%%%%%%%%%
\begin{figure*}[ht]
\begin{center}
\epsfig{file=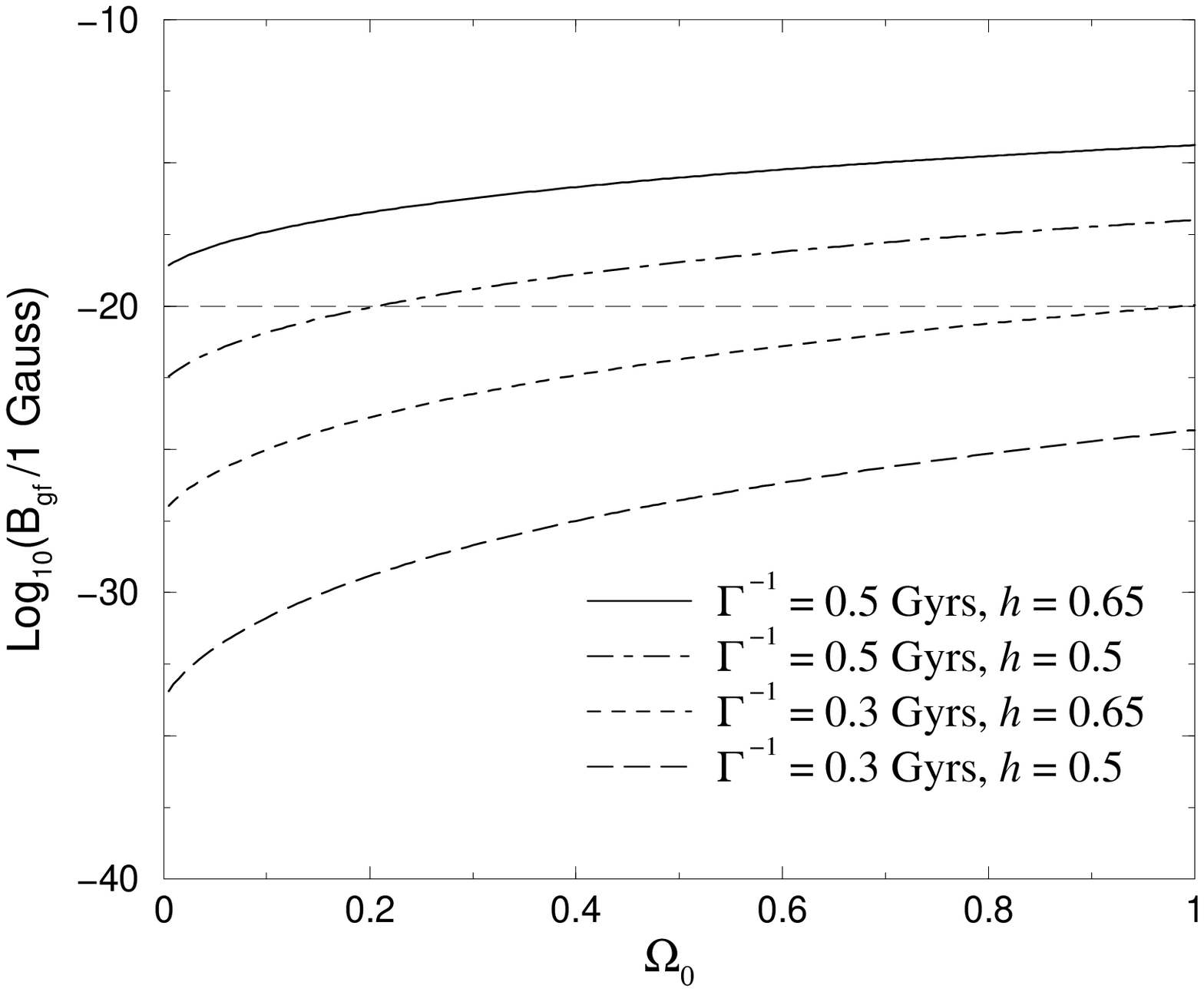 ,width=7.8cm}\hspace*{5mm}
\epsfig{file=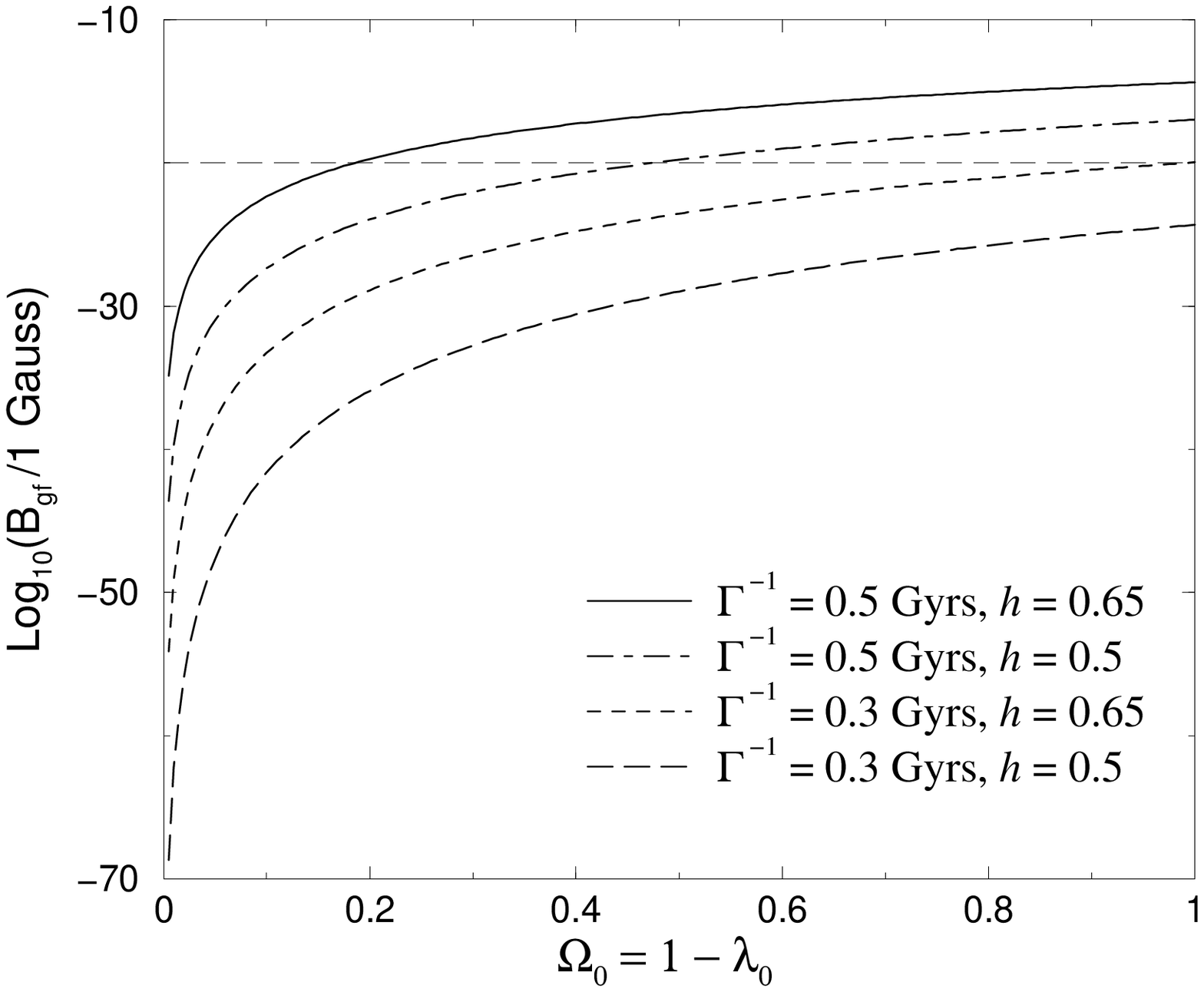 ,width=7.8cm}
\end{center}
\vspace*{-6.5cm}\noindent
\hspace*{2.2cm}{\large a)} \hspace*{7.85cm}{\large b)}
\vspace*{5.4cm}
\lbfig{gfz=0}
\begin{center}
\caption{Lower bound on the seed field at galaxy formation
$B_{\rm gf}$ vs $\Omega_0$:
(a) universe with $\Lambda=0$, (b) flat $\Lambda$ universe.}
\end{center}
\end{figure*}
%%%%%%%%%%%%%%%%%%%%%%%%%%%%%%%%%%%%%%%%%%%%%%%%%%%%%%%%%%%%%%%%%%%%%%

This is the result that we have sought. It shows that the relationship
between the average density of a galaxy $\bar{\rho}_{\rm gal}$
and the time of completed galaxy formation $t_{\rm gf}$ is independent
of cosmology.
The current
galactic density is a quantity which can be measured using methods
that also do not depend on cosmology.
The redshift of galaxy formation $z_{\rm gf}$
corresponding
to a given value of
$\bar{\rho}_{\rm gal}$ is given implicitly by well-known
time-redshift relations (see e.g. Refs.\ \cite{kolb} and \cite{lahav}).
It may at first
seem mysterious that $\Omega_0$
does not enter in Eq.~(\ref{rhobar}) or any of the derivations leading
to it. The reason is that (for $\Lambda=0$)
the same average local density
(larger than the critical density) is required for a spherical region
to
collapse regardless of the density of the surrounding universe.
By Birkhoff's theorem the evolution therefore
proceeds in an identical manner.

We are now in a position to calculate bounds on magnetic seed fields
$B_{\rm gf}$ at the time of completed galaxy formation in different
cosmologies. We take
$B_0=10^{-6}$ G and $\bar{\rho}_{\rm gal}=10^{-24}$
g$\,$cm$^{-3}$. The latter value corresponds to the average
density of the galactic halo rather than
the central disc, whose density
is $\sim 10^{-23}$ g$\,$cm$^{-3}$. The reason for this
choice is that the spherical collapse model uses the simplified
assumption of a
uniform-density ``top-hat'' profile of the galactic density distribution,
and since the halo comprises most of the volume of the
galaxy, this value seems more appropriate. The precise value of
 $\bar{\rho}_{\rm gal}$ is of little
importance
as our results
are quite insensitive to it.

The results are displayed in Fig.~\ref{gfz=0}(a) for a
$\Lambda=0$ universe and
in Fig.~\ref{gfz=0}(b)
for a flat $\Lambda$ universe ($\Omega_0+\lambda_0=1$).
The quantity $h$ is the Hubble parameter $H_0$
in units of $100$ km sec$^{-1}$ Mpc$^{-1}$.
For comparison, the straight horizontal line in each plot shows the
constraint of $B_{\rm gf}\geq$ \mbox{$10^{-20}$ G}
given by Rees \cite{rees}.
It can be seen on these graphs that in an open universe, and
particularly
in a universe with a significant cosmological constant $\Lambda$, this
requirement is too strong. For reasonable cosmological parameters
and the same value of $\Gamma$,
the dynamo mechanism could generate currently observed galactic
magnetic
fields from a seed field
of the order of $10^{-30}$ G or less at the
completion of galaxy formation provided that the seed field
is coherent on a scale
$\xi_{\rm gf}\gsim 100$ pc.

We shall now evolve these bounds back to
the time of radiation decoupling, taking the conservative view
that there is no magnetohydrodynamic turbulence or
dynamo mechanism operating during gravitational
collapse (see, however, Ref.~\cite{Kulsrud} for
more optimistic proposals). The magnetic field is assumed to be
frozen into
the plasma and its evolution is determined by flux conservation
$Br^2= {\rm const.}$,
where $r$ is a length scale evolving with the matter, i.e.\
$r\sim (\bar{\rho})^{-1/3}$.
Care must be taken not to associate this length scale with the
scale factor $a(t)$, as a collapsing galaxy is decoupled from the
Hubble expansion. One obtains \cite{turner}
\begin{equation} \label{Bdec}
{B_{\rm gf} \over B_{\rm dec}} = \left({\bar{\rho}_{\rm gal} \over
  \rho_{\rm dec}}\right)^{\!\!2 / 3}\!\! =
 \left({\bar{\rho}_{\rm gal} \over
  \rho_{0}}\right)^{\!\!2 / 3}\!\! {1 \over (1+z_{\rm dec})^2}~,
\end{equation}
where we have used the energy conservation relation
$\rho = \rho_0 (1+z)^3$ for the matter component
 and
the fact that the matter density
$\rho_{\rm dec}$ at the
epoch
$t_{\rm dec}$ is
very nearly
uniform.
The redshift of
radiation decoupling, $z_{\rm dec}$, has a weak dependence on the product
$\Omega_B h^2$ involving the fraction of critical density in baryons
$\Omega_B$ and is constrained to lie in the interval
$1200\lsim z_{\rm dec} \lsim 1400$
\cite{kolb}.

Note that the magnetic field will {\em decrease\/} between $t_{\rm
dec}$
and $t_{\rm gf}$, since by virtue of the Hubble expansion
the physical volume of the galaxy is larger than the volume containing
the same mass at $t_{\rm dec}$. The depletion depends on the
cosmological parameters via
the present matter density $\rho_0\approx 1.88\times 10^{-29}
\Omega_0 h^2$ [g cm$^{-3}$]. It can be seen that the depletion is somewhat
smaller in universes with $\Omega_0 < 1$. This further increases the
effect of cosmological parameters in relaxing
the bounds on primordial seed fields. The resulting bounds for
$B_{\rm dec}$ are shown in Fig.~\ref{decz=0}(a) and Fig.~\ref{decz=0}(b)
for a $\Lambda=0$ universe and for
a flat $\Lambda$ universe, respectively.

%%%%%%%%%%%%%%%%%%%%%%%%%%%%%%%%%%%%%%%%%%%%%%%%%%%%%%%%%%%%%%%%%%%%%%
\begin{figure*}[ht]
\begin{center}
\epsfig{file=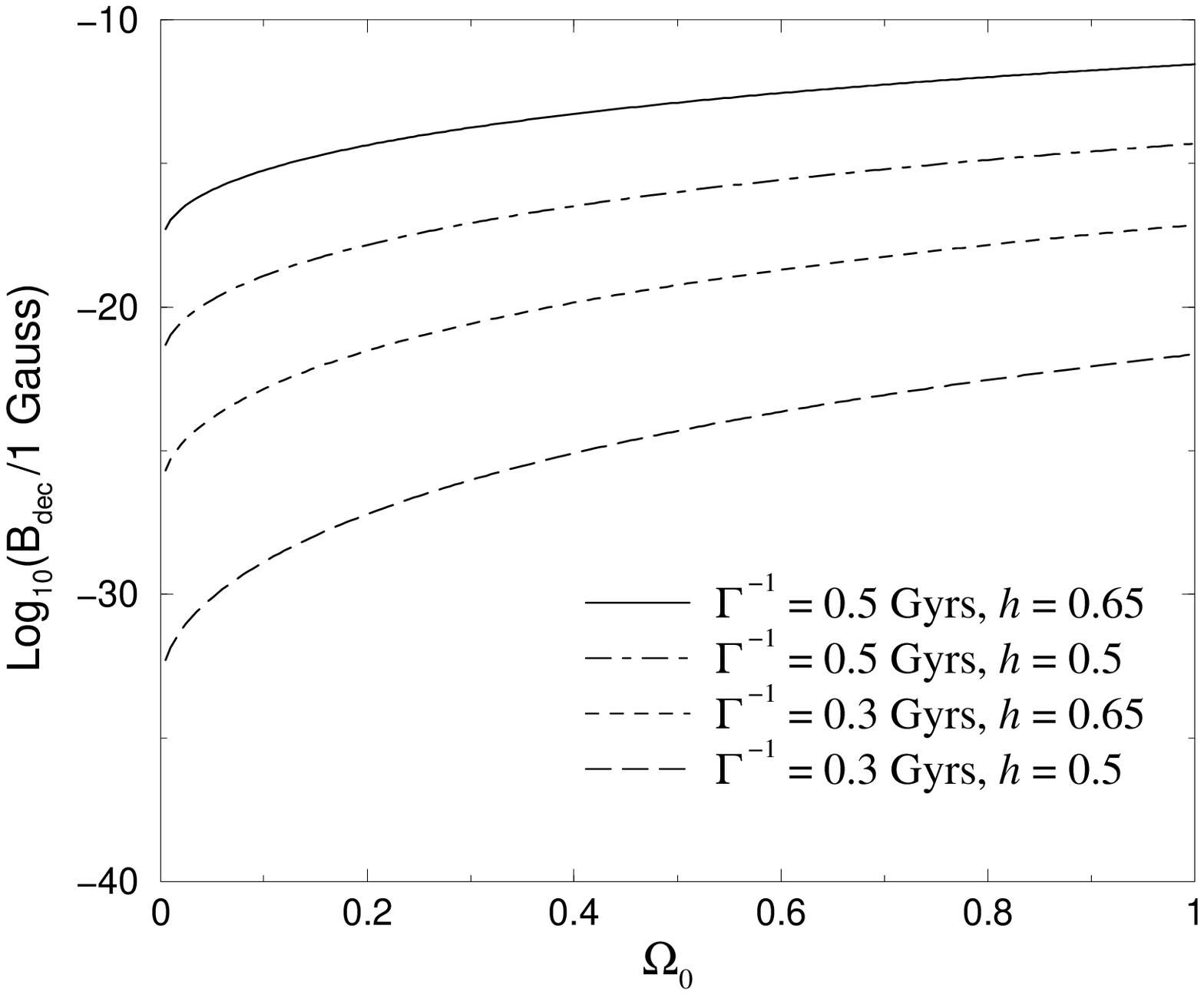 ,width=7.8cm}\hspace*{5mm}
\epsfig{file=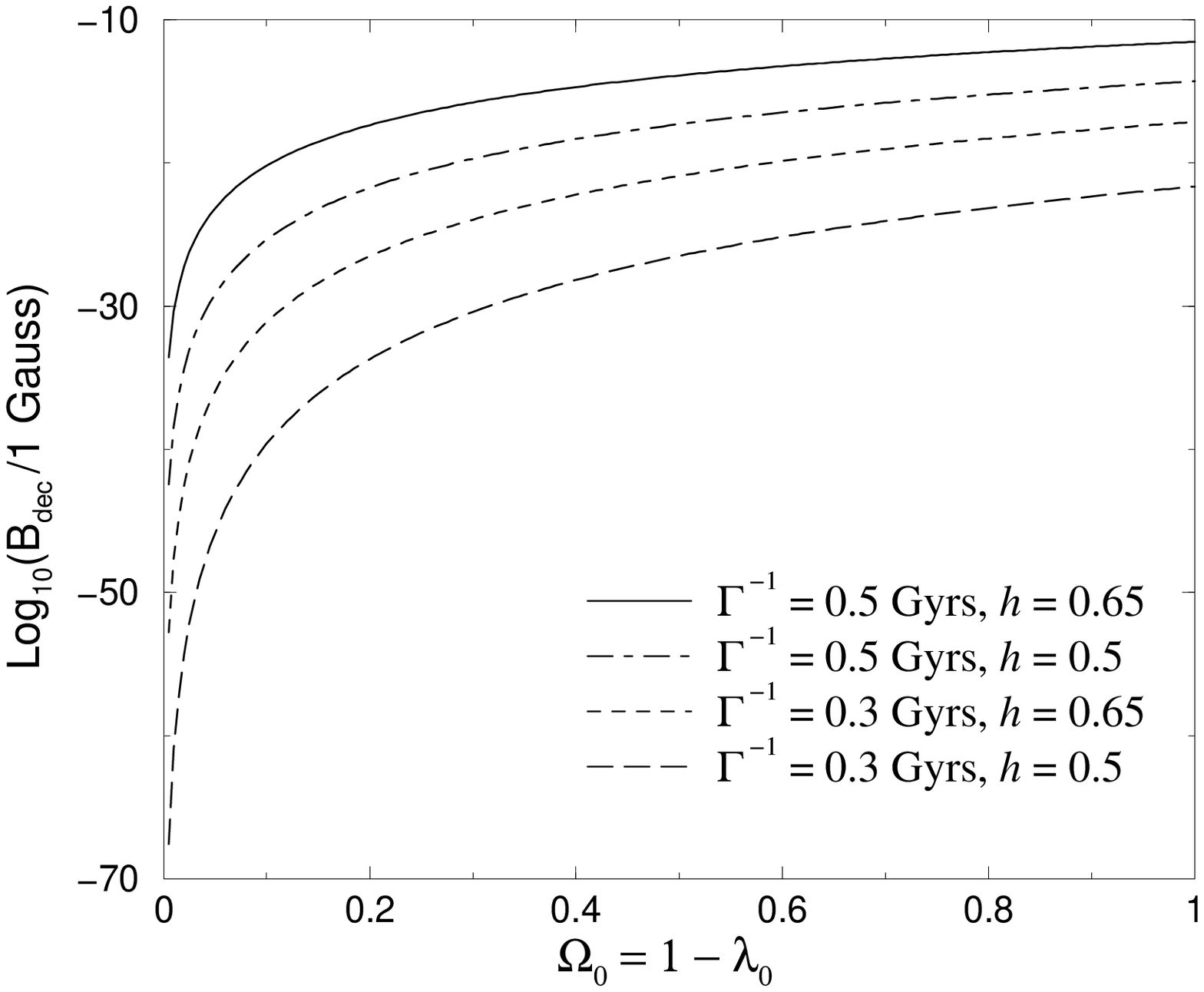 ,width=7.8cm}
\end{center}
\vspace*{-6.5cm}\noindent
\hspace*{2.2cm}{\large a)} \hspace*{7.85cm}{\large b)}
\vspace*{5.4cm}
\lbfig{decz=0}
\begin{center}
\caption{Lower bound on the seed field at radiation decoupling
$B_{\rm dec}$ vs $\Omega_0$:
(a) universe with $\Lambda=0$, (b) flat $\Lambda$ universe.}
\end{center}
\end{figure*}
%%%%%%%%%%%%%%%%%%%%%%%%%%%%%%%%%%%%%%%%%%%%%%%%%%%%%%%%%%%%%%%%%%%%%%

We shall now address the issue of the correlation length of the
magnetic field. In order for the galactic dynamo to begin to
operate, the
correlation length of the seed field at the time of completed galaxy
formation must satisfy $\xi_{\rm gf} \gsim 100$ pc \cite{dynamo}.
\footnote{A more conservative bound, used by many
authors, is $\xi_{\rm gf} \gsim 1$ kpc.}
Using the spherical collapse model, one can calculate the physical
scale $r_{\rm dec}$ at the time of radiation decoupling that will evolve
into the size of a galaxy. At any time before
 the onset of gravitational collapse
 the matter density follows the Hubble expansion and it makes sense
to express $r_{\rm dec}$ in the constant comoving quantity $x$ defined by
$r=a(t) x$. The comoving scale $x$ corresponding to a galaxy is given
by \cite{padmanabhan}
\begin{equation} \label{comoving2}
x_{\rm gal} =
0.95 \left(\Omega_0 h^2\right)^{-1/3} M_{12}^{1/3}\  [{\rm Mpc}]~,
\end{equation}
where $M_{12}=M/10^{12} M_{\odot}$.

The correlation length $\xi$ can be written as a fraction
of the radius of the galaxy,
 $\xi = \eta\, r_{\rm vir}$. With the simplified assumption
of the spherical collapse model that the collapsing region has
uniform density, the collapse is homogeneous and isotropic and
different scales collapse proportionately. Assuming that the magnetic field
is frozen into the plasma between $t_{\rm dec}$ and $t_{\rm gf}$ we have
$x_{\rm corr} = \eta x_{\rm gal}$. For a galaxy, $M_{12}\approx 0.1$, and
the typical length scale of the
turbulent motion, $\xi_{\rm turb}=100$ pc,
corresponds to $\eta\approx 1/150$, giving
the following bound on the comoving correlation length
\begin{equation}
\label{xcorr}
x_{\rm corr}\gsim x_{\rm turb\/} = 5~\mbox{--}~10~{\rm kpc}~
\end{equation}
for observationally realistic values $0.25>\Omega_0 h^2>0.025$.
This bound is somewhat higher than that stated in
Ref.~\cite{kostas}.
The bound should not be applied
before $t_{\rm dec}$, since the correlation length
then evolves according to
complicated magnetohydrodynamic processes and is not
proportional to
the scale factor $a(t)$ \cite{kostas}.

In general, primordial seed fields produced by particle-physics or
field-theory mechanisms are too incoherent to meet the requirement posed
by Eq.~(\ref{xcorr}).
However,
there is a possibility even for a less correlated magnetic field
to pass the
requirement provided that it has sufficient strength to satisfy the
bound on $B_{\rm dec}$
after r.m.s.\ coarse-graining over the
scale given by $x_{\rm turb}$. The said procedure results in an
root-mean-square field
\begin{equation}
B_{\rm rms} =
\left( \frac{x_{\rm dec}}{x_{\rm turb}}\right)^{d/2}\!\!
B_{\rm dec}~,
\end{equation}
where $B_{\rm rms}$ is the quantity that
must satisfy the bound given in Fig.~\ref{decz=0},
with  $B_{\rm dec}$ and $x_{\rm dec}$ being
the strength and comoving correlation length, respectively,
of the primordial
seed field
evolved from formation to $t_{\rm dec}$. The exponent $d$ can equal 1, 2,
or 3 depending on
the averaging procedure used. This complicated issue \cite{average}
shall not be addressed in this paper.

There have been observations of micro-Gauss fields at redshifts of
$z=0.395$ \cite{kronberg} and $z=2$
\cite{wolfe}, although the latter has
been criticised \cite{perry}. If correct, these observations
are difficult to explain in a flat universe with $\Lambda=0$. They
may,
however, be easier to understand in an open or $\Lambda$ universe.
Applying our model to the $z=0.395$ case, with $B_{0.395}=10^{-6}$ G,
we obtain for a flat $\Lambda$ universe
the bounds
at $t=t_{\rm dec}$ shown in
Fig.~\ref{highzfig}. Hence a seed field of $10^{-20}$ G
at $t_{\rm dec}$,
or equivalently $10^{-23}$ G at $t_{\rm gf}$,
could account for this observation.

%%%%%%%%%%%%%%%%%%%%%%%%%%%%%%%%%%%%%%%%%%%%%%%%%%%%%%%%%%%%%%%%%%%%%%
\begin{figure}[htb]
\begin{center}
\epsfig{file=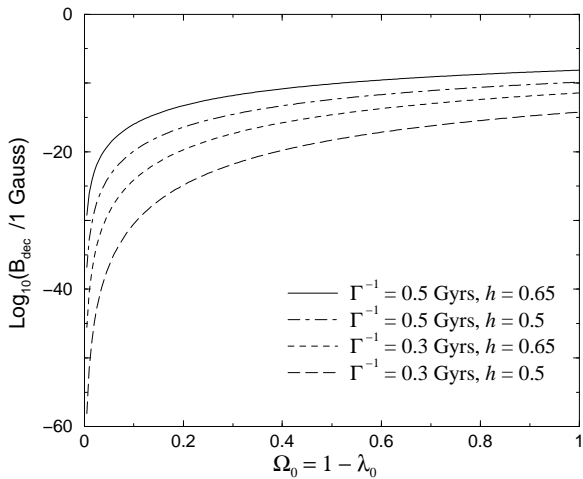 ,width=7.8cm}
\lbfig{highzfig}
\caption{Lower bound on $B_{\rm dec}$ vs $\Omega_0$ for generating
a field strength of $10^{-6}$ G at
redshift $z=0.395$ by the dynamo mechanism in a universe with
$\lambda_0+\Omega_0=1$.}
\end{center}
\end{figure}
%%%%%%%%%%%%%%%%%%%%%%%%%%%%%%%%%%%%%%%%%%%%%%%%%%%%%%%%%%%%%%%%%%%%%%

If we attempt a similar analysis in the $z=2$ case, the required
seed field is sufficiently high that it would have other cosmological
implications, e.g. on the CMB
\cite{Rubinstein} or structure formation \cite{rosner}.
Consequently we conclude that, unless the dynamo parameters are
radically different for high column density Ly-$\alpha$ clouds
(e.g. if they have fast-spinning cores
and thereby have a higher angular-velocity gradient $|r\,d\omega/dr|$
\cite{fastcore})
these observations cannot be explained by amplification of a
primordial
seed field by a galactic dynamo.

In this paper we have reconsidered the constraints on the primordial magnetic
field required to seed the galactic dynamo in the light of
recent cosmological advances. We have shown that, in an open
universe or a universe with $\Lambda > 0$, a much smaller seed field
is required to explain the observed micro-Gauss fields in galaxies.
As a consequence, mechanisms of
primordial magnetic seed-field generation that
had previously been ruled out, on the grounds of giving too small
strength or correlation length, could well be viable.
We have evolved the bounds back to
the epoch of radiation decoupling $t_{\rm
dec}$,
assuming that
from
$t_{\rm dec}$ to the present the magnetic
field is frozen into the plasma and
evolves first via flux
conservation and thereafter by amplification via a dynamo mechanism.
The remaining problem is to
evolve primordial magnetic fields from the time of their generation
to $t_{\rm dec}$ taking into account various plasma effects. A
step in this direction  has been taken in Ref.~\cite{kostas}. This
work needs to be generalised to different cosmologies, although it can
be expected that the main cosmological effects occur at late times.

We are grateful to K.\ Dimopoulos, L.\ Hui, M.J.\  Rees and N.\ Weiss
for helpful discussions. This work was supported in part by the U.K.\ PPARC.
Support for M.L.\ was provided by a PPARC studentship and by
Fitzwilliam College, Cambridge;
for O.T.\ by
the European Commission's TMR programme under Contract
No.~ERBFMBI-CT97-2697.
\vspace*{-5mm}
\setlength{\parskip}{-15mm}


\begin{thebibliography}{99}
\setlength{\topsep}{0pt}
\setlength{\parskip}{0pt}
\bibitem{obsreview}P.P.\ Kronberg, Rep.\ Prog.\ Phys.\ {\bf 57}, 325 (1994);
 R.\ Beck, A.\ Brandenburg, D.\ Moss, A.\ Shukurov,
and D.\ Sokoloff, Annu.\ Rev.\ Astron.\ Astrophys.\ {\bf 34},
155 (1996).

\bibitem{dynamo} Ya.B.\ Zeldovich, A.A.\ Ruzmaikin and D.D.\ Sokolov,
{\em Magnetic Fields in Astrophysics\/} (Gordon and Breach, New York,
1983).

\bibitem{moffatt} H.K.\ Moffatt, {\em Magnetic Field Generation in
    Electrically Conducting Fluids\/} (Cambridge University Press,
  Cambridge, 1978); E.\ N.\ Parker,
{\em Cosmological Magnetic Fields\/} (Clarendon,
Oxford, 1979).

\bibitem{Kulsrud}  H.\ Lesch and M.\ Chiba,
Astron.\ Astrophys.\ {\bf 297}, 305L (1995);
R.M.~Kulsrud, R.\ Cen, J.P.\ Ostriker and D.\ Ryu,
Ap.\ J.\ {\bf 480}, 481 (1997).

%\bibitem{Biermann} L.\ Biermann, Zs.\ Naturforsch.\ {\bf 5a}, 65
%(1950).

\bibitem{field} Ya.\ B.\ Zeldovich, Zh.\ Eksp.\ Teor.\ Fiz.\ {\bf 48},
986 (1964) [Sov.\ Phys.\ JETP {\bf 21}, 656 (1965)];
K.\ Thorne, Bull.\ Am.\ Phys.\ Soc.\ {\bf 11}, 340 (1966),
Ap.\ J.\ {\bf 148}, 51 (1967).

\bibitem{thewholebunchincludingOlasOriginpaper} For a review and
further references
see, e.g., K.\ Enqvist, Int.\ J.\ Mod.\ Phys.\ {\bf D7} 331 (1998);
A.\ Olinto, astro-ph/9807051;  O. T\"{o}rnkvist,
Phys.\ Rev.\ {\bf D58}, 043501 (1998).
%hep-ph/9707513

\bibitem{KibVil} T.W.B.\ Kibble, A.\ Vilenkin, {\em Phys.\ Rev.\ }{\bf D52},
679 (1995); O. T\"{o}rnkvist, hep-ph/9902432.
%E.J.\ Copeland, P.M.\ Saffin and O. T\"{o}rnkvist, in preparation.

\bibitem{turner} M.S.\ Turner and L.M.\ Widrow, Phys.\ Rev.\ {\bf D37},
  2743 (1988).

\bibitem{davis} A.-C.\ Davis and K.\ Dimopoulos, Phys.\ Rev.\ {\bf
D55}, 7398 (1997).


\bibitem{supernovae} S.\ Perlmutter et al.,
%{\em Measurements of
%Omega and Lambda from 42 High-Redshift Supernovae\/},
astro-ph 9812133, to appear
in Ap.\ J.;
P.M.\ Garnavich et al., Ap.\ J.\ {\bf 509}, 74G (1998).

\bibitem{CMB} M.\ Tegmark, Ap.\ J.\ {\bf 514L}, 69T (1999);
G.\ Efstathiou, S.\ L.\ Bridle, A.\ N.\ Lasenby and M.\ P.\ Hobson,
%{\em Constraints on $\Omega_\Lambda$ and $\Omega_m$ from Distant
%Type 1a Supernovae and Cosmic Microwave
%     Background Anisotropies\/},
astro-ph/9812226, submitted to
Mon.\ Not.\ R.\ Astr.\ Soc.\

\bibitem{kostas} K.\ Dimopoulos and A.C.\ Davis,
Phys.\  Lett.\ {\bf B390}  87 (1997);
K.\ Dimopoulos, Ph.D.\ Thesis, University of Cambridge
(1997).


\bibitem{RST} A.A.\ Ruzmaikin, D.D.\ Sokolov and V.I.\ Turchaninov,
  Astron.\ Zh.\ {\bf 57}, 311 (1980) [Sov.\ Astron.\ {\bf 24},
182 (1980)].

\bibitem{kolb} E.\ W.\ Kolb and M.\ S.\ Turner, {\em The Early Universe}
  (Addison-Wesley, 1990).

\bibitem{lahav} O.\ Lahav, P.B.\ Lilje, J.R.\ Primack and M.J.\ Rees,
  Mon.\ Not.\ R.\ Astr.\ Soc.\ {\bf 251}, 128 (1991).

\bibitem{peebles} P.J.E.\ Peebles, {\em Large Scale Structure of the
    Universe\/} (Princeton University Press, Princeton, 1980).

\bibitem{padmanabhan} T.\ Padmanabhan, {\em Structure Formation in the
    Universe\/} (Cambridge University Press, Cambridge, 1993).

\bibitem{edwards} D.\ Edwards, Mon.\ Not.\ R.\ Astr.\ Soc.\ {\bf
    159}, 51 (1972).

\bibitem{rees} M.J.\ Rees, Q.\ Jl R.\ Astr.\ Soc.\ {\bf 28}, 197
  (1987).

\bibitem{average}  K.\ Enqvist and P.\ Olesen,
 Phys.\ Lett.\ {\bf B319}, 178 (1993); M.\ Hindmarsh and A.\ Everett,
Phys.\ Rev.\ {\bf D58}, 103505 (1998).

\bibitem{kronberg} P.P.\ Kronberg and J.J.\ Perry, Ap.\ J.\ {\bf
263}, 518 (1982); P.P.\ Kronberg, J.J.\ Perry and
  E.L.H.\ Zukowski, Ap.\ J. {\bf 387}, 528 (1992).

\bibitem{wolfe} A.M.\ Wolfe, K.M.\ Lanzetta and A.L.\ Oren, Ap.\
J. {\bf 388}, 17 (1992).

\bibitem{perry} J.J.\ Perry, A.M.\ Watson and P.P.\ Kronberg,
Ap.\ J.\ {\bf 406}, 407 (1993).

\bibitem{Rubinstein}
J.\ Magueijo, Phys.\ Rev.\ {\bf D49}, 671 (1994);
J.\ Adams, U.H.\ Danielsson, D.\ Grasso, and H.\ Rubinstein,
Phys.\ Lett.\ {\bf B388}, 253 (1996);
J.D.\ Barrow, P.G.\ Ferreira and J.\ Silk,
Phys.\ Rev.\ Lett.\ {\bf 78}, 3610 (1997).

\bibitem{rosner} I.\ Wasserman, Ap.\ J.\ {\bf 224}, 337 (1978);
E.\ Kim, A.V.\ Olinto and R.\ Rosner, Ap.\ J.\ {\bf 467}, 28K (1996).

\bibitem{fastcore} G.S.\ Bisnovatyi-Kogan, A.A.\ Ruzmaikin and
R.A.\ Syunyaev, Astron.\ Zh.\ {\bf 50}, 210 (1973) [Sov. Astron. {\bf
17}, 137 (1973)];
M.J.\ Rees, private communication.

\end{thebibliography}
\end{document}